\documentstyle[galley]{mn}
\onecolumn
\input{epsf}
\def\n{{\noindent}}

\title[ Statistics of Collapsed Objects]
{From Snakes to Stars, the Statistics of Collapsed Objects - I.
Lower--order Clustering Properties.\\}

\author[D. Munshi et al.]{Dipak Munshi$^{1,4}$,
Peter Coles$^{1,2}$ and Adrian L. Melott$^3$,\\ $^1$Astronomy
Unit,  Queen Mary and Westfield College, University of London,
London E1 4NS, UK\\ $^2$School of Physics \& Astronomy, University
of Nottingham, University Park, Nottingham NG7 2RD, UK
\\
 $^3$Department of Physics and Astronomy,
University of Kansas, Lawrence, Kansas 66045, USA \\
$^4$International School for Advanced Studies (SISSA),
Via Beirut 2-4, I-34013 Trieste, Italy}

\begin{document}

\maketitle

\begin{abstract}
The highly nonlinear regime of gravitational clustering  is
characterized by the presence of scale-invariance in the hierarchy
of many-body correlation functions. Although the exact nature of
this correlation hierarchy can only be obtained by solving the
full set of BBGKY equations, useful insights can be obtained by
investigating the consequences of a generic scaling {\em ansatz}.
Extending earlier studies by Bernardeau \& Schaeffer (1992) we calculate the
detailed consequences of such scaling for the
implied behaviour of a number of statistical descriptors,
including some new ones, developed to provide useful diagnostics
of scale-invariance. We generalise  the two-point cumulant
correlators (now familiar to the literature) to a hierarchy of
multi-point cumulant correlators (MCC) and introduce the concept
of reduced cumulant correlators (RCC) and their related generating
functions. The description of these quantities in diagrammatic
form is particularly attractive. We show that every new vertex of
the tree representation of higher-order correlations has its own
reduced cumulant associated with it and, in the limit of large
separations, MCCs of arbitrary order can be expressed in terms of
RCCs of the same and lower order. Generating functions for these
RCCs are related to the generating functions of the underlying
tree vertices for matter distribution. Relating the generating
functions of RCCs with the statistics of {\em collapsed} objects
suggests a  scaling ansatz of a very general form for the
many-body correlation functions which, in turn, induces a similar
hierarchy for the correlation functions of overdense regions. In
this vein, we compute the lower-order $S_N$ parameters and
two-point cumulant correlators $C_{NM}$ for overdense regions and
study how they vary as a function of the initial power spectrum of
primordial density fluctuations. We also show that these results
match those obtained by the extended Press-Schechter formalism in
the limit of large mass.
\end{abstract}


\bigskip
\section{\bf Introduction}
\bigskip
\n
Gravity is scale-free. In the absence of any externally-imposed
length scale, such as might be set by initial conditions, it is
therefore reasonable to assume that gravitational clustering of
particles in the universe should evolve towards a scale-invariant
form, at least on small scales where gravitational effects
presumably dominate over initial conditions. Observations offer
support for such an idea, in that the observed two-point
correlation function $\xi(x)$ of galaxies is reasonably well
represented by a power law over quite a large range of length
scales,
\begin{equation}
\xi({\bf r}) = \Big ( {r \over 5h^{-1} {\rm Mpc}} \Big )^{-1.8}
\end{equation}
(Groth \& Peebles 1977; Davis \& Peebles 1983) for, say, $r$
between $100 h^{-1} {\rm kpc}$ and $ 10h^{-1}~{\rm~ Mpc}$.
Higher-order correlation functions of galaxies also appear to
satisfy a scale-invariant form, with $\xi_N \propto \xi_2^{N-1}$
as expected from the application of a general scaling ansatz
(Groth \& Peebles 1977; Fry \& Peebles 1978; Davis \& Peebles
1983; Szapudi et al. 1992). For example, the observed three point
function, $\xi_3$, is well-established to have a hierarchical form
\begin{equation}
\xi_3({\bf x}_a, {\bf x}_b, {\bf x}_c) =~ \xi_{abc}~ = Q(\xi_{ab}\xi_{bc} +
\xi_{ac}\xi_{ab} + \xi_{ac}\xi_{bc}),
\end{equation}
where $\xi_{ab}=\xi ({\bf x}_a, {\bf x}_b)$, etc. In a similar
spirit, the four-point correlation function can be expressed as a
combination of four-point graphs with two different topologies --
``snake'' and ``star'' -- with corresponding amplitudes $R_a$ and
$R_b$ respectively.
\begin{equation}
\xi_4({\bf x}_a, {\bf x}_b, {\bf x}_c, {\bf x}_d) = ~ \xi_{abcd}~ =
R_a(\xi_{ab}\xi_{bc}\xi_{cd}
+ \dots (\rm 12~terms)) + R_b(\xi_{ab}\xi_{ac}\xi_{ad}
+ \dots (\rm 4~terms))
\end{equation}
(Fry \& Peebles 1978, Fry 1984a, Bonometto et al. 1990, Valdarini \&
Borgani 1991). It is natural on the basis of these considerations
to infer that all N-point correlation functions can be expressed
as a sum over all possible N-tree graphs with (in general)
different amplitudes for each tree diagram. It is also generally
assumed that there is no shape-dependence in these tree amplitudes
in the highly non-linear regime:
\begin{equation}
\xi_N( {\bf r}_1, \dots {\bf r}_N ) = \sum_{\alpha, \rm N-trees}
Q_{N,\alpha} \sum_{\rm labellings} \prod_{\rm edges}^{(N-1)}
\xi({\bf r}_i, {\bf r}_j).
\end{equation}
Some numerical experiments have confirmed the validity of such an
assumption, at least as far as three-point correlation function
is concerned (Scoccimarro et al. 1998).

This tree-level model of hierarchical clustering however is a
particular case of a more general scaling ansatz proposed by
Balian \& Schaeffer (1988, 1989), in which the $N$--point
correlation functions can be written in the form
\begin{equation}
\xi_N( \lambda {\bf r}_1, \dots  \lambda {\bf r}_N ) =
\lambda^{-\gamma(N-1)} \xi_N( {\bf r}_1, \dots {\bf r}_N ).
\label{hierarchical}
\end{equation}
Rigorous theoretical motivation for these assumptions are,
however, yet to be forthcoming. Considerable progress has been
made recently in understanding the evolution of gravitational
clustering in limit of weak clustering, i.e. in the quasi-linear
regime (e.g. Sahni \& Coles 1995). In particular, it has been
shown that, in the quasi-linear regime, a tree-level hierarchy of
correlation functions develops which satisfies ${\bar \xi_N}
\propto {\bar \xi_2}^{N-1}$ at the limit of vanishing variance. On
the other hand, the intermediate regime, where loop corrections to
tree-level diagrams become important, and the highly non-linear
regime, where perturbative calculations break down completely,
remain poorly understood and answers to many questions are still
unclear.

Theoretical studies of the evolution of correlation functions in
the highly nonlinear regime have mainly concentrated on possible
closure schemes for the Born-Bogoliubov-Green-Kirkwood-Yvon
(BBGKY) equations (Davis \& Peebles 1977), stability properties of
these closure schemes (Ruamsuwan \& Fry 1992) and different
phase-space separation schemes for hierarchical solutions of
arbitrary order (Fry 1982; Hamilton 1988).  Although a general
solution of the BBGKY equations in nonlinear regime is still
lacking, various interesting characteristics of the count
probability distribution function (CPDF) and void probability
distribution function (VPF) have been
 established by Balian \& Schaeffer(1992). These predictions have
 also been tested against numerical simulations, in both two and
 three dimensions, and were found to be in good agreement.

However, most of the analytical results available so far relate to
quantities associated only with one-point cumulants. Examples are
the so-called $S_N$ parameters, as well as the CPDF and VPF
mentioned above. Multi-point cumulant correlators (MCC), which we
shall introduce in this paper, are a natural generalization of the
$S_N$ parameters and the two-point cumulant correlators (2CC)
which are generally used to analyze galaxy catalogs and compare
observational data with different models of structure formation.
Cumulant correlators were proposed recently by Szapudi \& Szalay
(1997) and shown to be very efficient for extracting clustering
information from galaxy surveys. In particular, they can be used
to test gravitational instability scenario and constrain the
initial power spectral index. Analytical predictions regarding the
2CCs were made by Bernardeau (1995) using tree-level perturbation
theory in the limit of large separation. In particular, he showed
how the {\em bias} of overdense cells with respect to the
underlying mass distribution is related to the generating function
of the 2CCs. These predictions were also successfully tested
against numerical simulations. Such calculations were also
extended for application to projected (angular) catalogues and
have now been shown to be in reasonable agreement with estimates
of these quantities extracted from the APM survey (Munshi, Melott
\& Coles 1998).

Bernardeau \& Schaeffer (1992) developed a systematic series
expansion for the multi-point void probability distribution
function (MVPF) which is essentially the generating function of
the multi-point count probability distribution function (MCPDF).
This expansion again applies only in the highly nonlinear regime.
Using such a series expansion, based on powers of $\xi_{ij}/ \bar
\xi_i$ where $\bar \xi_i$ is the variance in the $i$th cell, they
were able to evaluate the bias associated with overdense regions.
The bias thus calculated, which was analytically shown to depend
only on a scaling variable associated with the collapsed object
and an intrinsic property of the object, was later found to be in
very good agreement with numerical calculations. It is clearly an
interesting task to extend such analysis further and check
predictions against numerical simulations.

If one can believe that galaxies are in some way identified with
overdensities in the matter distribution then observed estimates
of the two-point cumulant correlators from a galaxy survey depend
upon the bias of such regions relative to the underlying mass
distribution. Multipoint cumulant correlators  allow more
information to be deduced about the properties of bias, because
the quantities related to the $S_N$ parameters of overdense cells.
Complete knowledge of all $S_N$ parameters (which are related to
the generating function of MCCs to arbitrary order) can also help
us in constructing the count probability distribution, and related
void probability distribution, for collapsed objects.

Unfortunately, there is at present no robust theoretical framework
other than the hierarchical ansatz within which such objects can
be studied. One-point statistics of collapsed objects have been
studied by Mo \& White (1996) using the extended Press-Schechter
formalism (Press \& Schechter 1974; Bond et al. 1991; Bower 1991;
Lacey \& Cole 1993; Kauffmann \& White 1993) but it remains to be
seen whether this formulation can be used to derive multi-point
statistics for collapsed objects. The same is largely true of
``peaks theory'' which is based on an argument used by Kaiser
(1984)  to explain the strong clustering of Abell clusters using
the properties of high-density regions in Gaussian fields. This
formalism was subsequently developed by Bardeen et al.(1986) and
its use in cosmology is now widespread. However, there is evidence
that correspondence between dark haloes and density peaks is not
particularly good (Frenk et al. 1998; Katz, Quinn \& Gelb 1993)
and one needs to take the gravitationally-induced motions of peaks
into account to compute the final clustering properties of halos.
Progress has recently been made in overcoming these difficulties
by Bond \& Myers (1996a,b)  in the ``Peak-Patch Formalism'' at the
expense of considerable technical complexity. Other efforts to
study statistics of collapsed objects include use of the
Zel'dovich approximation in nonlinear regime after suitably
smoothing the initial potential and using the statistics of
singularities of different types (Catelan et al. 1998; Lee \&
Shandarin 1997, 1998).

It is also important to note that earlier derivation of bias and
$S_3$ for collapsed objects in the context of hierarchical ansatz
were done within the context of certain specific approximations
 and it is interesting to check their validity  by extending such
 studies to higher orders,
as they are widely used for error estimations of measurements from
galaxy catalogs (Szapudi \& Colombi 1996), as well as the
derivation of mass function of collapsed objects (Valageas \&
Schaeffer 1997).

In a completely different context it was shown by Colombi et al.
(1996) that normalized one point cumulants or $S_N$ parameters for
underlying  mass distribution do not become constant in highly
non-linear regime, in direct contrast to hierarchal ansatz. Which
is also connected to the question, if gravitational clustering at
highly non-linear regime do satisfy the stable clustering ansatz
(Peebles, 1980). Connection between stable clustering and
hierarchical ansatz still remains  unclear and it is also related
to the question, if different initial conditions do lead to
similar halo profiles. In a recent study,  using stability
analysis of nonlinear clustering   Yano \& Gouda (1998) showed
that stable clustering may not be the most natural outcome of
gravitational clustering and departure from stable clustering
ansatz may mean departure from hierarchical ansatz. However their
results depend on specific closure schemes. In any case such
issues can only be tested if different prediction regarding
hierarchical ansatz are derived analytically and tested against
numerical simulations to find out their limitations, which is one
of the motivation behind present study.

The paper is organized as follows. In next section we develop a diagrammatic
method to represent cumulant correlators of arbitrary order. Using these rules
 it is also possible to compute these quantities to arbitrary order.
In section $\S 3$ we use the method of Bernardeau \& Schaeffer
(1992) to compute generating functions of RCCs to $5^{th}$ order.
In section $\S 4$ we use results of section $\S 3$ to compute
$S_N$ parameters for collapsed objects. We discuss these results
in the last section, and place them in the context of other
developments in this field.

\section{Multi-point Cumulant correlators and their Generating Functions}
Higher-order reduced correlation functions are typically dominated
by contributions from high peaks of the density fields. It is
reasonable, therefore, to infer that the generating functions  of
MCCs are related in some way to the statistics of collapsed
objects. According to the hierarchical ansatz, those tree-level
diagrams which contribute to reduced correlation function of the
form $\langle \delta({\bf x}_a)^p \dots \delta({\bf x}_h)^w
\rangle$ are of order $p+ \dots +w$. MCCs can be constructed by
identifying $p$ position arguments as ${\bf x}_a$, $q$ arguments
as ${\bf x}_b$, and so on, with the ${\bf x}_a$ representing the
positions of ``cells'' defined by some implicit smoothing of the
density field. Not all contributions constructed in this way are
of same order of magnitude. The dominant contribution will always
come from those terms which have the minimum number of external
lines joining different cells, since every extra line joining
different cells contributes an extra factor of $\xi_{ij}/ \bar
\xi_i << 1$ to the overall amplitude of the diagram.

With such a construction, the complete tree diagram can be
separated in two parts: (i) external lines joining different
cells; and (ii) internal lines joining the different vertices in
the same cell in such a way as to keep the external tree topology
unchanged. The number of internal degrees of freedom of the
diagram corresponds to the number of different ways the internal
vertices in a particular node can be arranged without altering the
external connectivity of the tree. The external tree configuration
connecting different nodes representing different cells on the
other hand is very similar to the underlying tree structure of
matter correlation function and depends only on the number of
nodes in the tree. The number of different topologies for external
tree increases with the number of nodes.

The difference between the external tree structure connecting
different nodes and the underlying tree structure of the matter
correlations connecting different vertices is that nodes in the
tree representation of MCCs carry an internal degree of freedom
due to the internal tree structure and represent an entire cell,
whereas for the underlying matter distribution, each vertex
represents a single point in space and does not carry any internal
structure. If we denote by $C_{p1}$ the number ways $p$ internal
points can be arranged with one external leg remaining to be
connected with another cell we can express the 2CCs as
\begin{equation}
\langle \delta^p({\bf x}_a) \delta^q({\bf x}_b) \rangle =
[C_{p1}\xi_{ab}C_{q1} ] {\bar \xi}^{p + q -2} = C_{pq} \xi_{ab} {\bar
\xi}^{p + q - 2}
\end{equation}
Similarly, we can define $C_{p11}$ as an internal degree of
freedom for a node with $p$ internal vertices and $2$ external
legs. The 3CCs now can be expressed in terms $C_{p11}$ and
$C_{p1}$. In practice $C_{p11}$ can be described as number of
different possible ways in which $p$ internal vertices can be
rearranged keeping two external legs free:
\begin{equation}
\langle \delta^p({\bf x}_a) \delta^q({\bf x}_b) \delta^r({\bf x}_c) \rangle =
\big [ C_{q1} \xi_{ab} C_{p11} \xi_{ac} C_{r1} + C_{p1} \xi_{ab} C_{q11} \xi_{bc}
C_{r1} + C_{p1} \xi_{ac} C_{r11} \xi_{bc} C_{p1} \big ] {\bar \xi}^{p + q +r -3}
= C_{pqr} \xi_{abc} {\bar \xi}^{p + q + r - 3}
\end{equation}
A similar decomposition of $4CCs$ and $5CCs$ can be used to define
$C_{p111}$ and $C_{p1111}$:
\begin{eqnarray}
\langle \delta^p({\bf x}_a) \delta^q({\bf x}_b) \delta^r({\bf
x}_c) \delta^s({\bf x}_d) \rangle & = & \big [ C_{q1} \xi_{ab}
C_{p11} \xi_{ad} C_{s1} \xi_{ac} C_{r1} + ({\rm
cyclic~permutations}) \nonumber \\
 & & +  ~C_{q1}\xi_{ab}C_{p11}\xi_{ac}C_{r11} \xi_{cd} C_{s1} +
 ({\rm cyclic~permutations}) \big ]{\bar \xi}^{p + q +r + s -4} \nonumber
\\
& = & C_{pqrs} \xi_{abcd} {\bar \xi}^{p + q + r + s - 4} ;
\end{eqnarray}
and
\begin{eqnarray}
\langle \delta^p({\bf x}_a) \delta^q({\bf x}_b) \delta^r({\bf
x}_c) \delta^s({\bf x}_d) \delta^t({\bf x}_e) \rangle & = & \big [
C_{p1111}\xi_{ab}C_{q1}\xi_{ac}C_{r1}\xi_{ae}C_{t1} + ({\rm
cyclic~permutations}) \nonumber\\  & & +
~C_{p111}\xi_{ad}C_{q11}\xi_{bc}C_{r1}\xi_{ad}C_{t1}\xi_{ae}C_{s1}
 + ({\rm cyclic~permutations})  \nonumber\\
& & +~ C_{p1} \xi_{ab} C_{q11} \xi_{bc} C_{r11} \xi_{cd} C_{s11}
\xi_{de} C_{t1} + ({\rm cyclic~permutations}) \big ] {\bar \xi}^{p
+ q +r + s +t -5}\nonumber \\  & = & C_{pqrst} \xi_{abcde} {\bar
\xi}^{p + q + r + s + t - 5} .
\end{eqnarray}

It is also possible to consider this set of new parameters $C_{p1\ldots1}$
describing the internal degrees of freedom as a set of reduced
cumulant correlators (RCCs) which can be used to decompose MCCs
$C_{p\ldots v}$. The number of subscript $1$s
that appear in the RCCs $C_{p1\ldots1}$ thus defined represents
the number of external trees and $p$ represents the number of
internal vertices. The total contribution from a cell is equal to
product of all vertices in the internal tree structure and  of the
number of different ways one can rearrange the trees keeping the
number of internal vertices and external legs fixed. The highest
order vertex that appears in an internal tree of $C_{p1}$ is $p$,
in $C_{p11}$ it is $p+1$ and so on. For the purposes of counting
different possible arrangements, the external hands are to be
considered distinguishable, as they link different external nodes
and hence represent one particular topology of the external tree.
Internal lines joining different vertices carry an weight of $\bar
\xi_i$ where, as above, $\bar \xi_i$ is the variance in the $i$th
cell. External lines carry an weight of $\xi_{ij}$, where
$\xi_{ij}$ represents the magnitude of the correlation function
joining $i$th and $j$th cells. Using these rules one can evaluate
RCCs to any order. We list here the contributions from different
topologies at lower orders:
\begin{eqnarray}
& & C_{11} =  1, C_{21} = 2\nu_2, C^{(a)}_{31} = 6 \nu_2^2,
C^{(b)}_{31}  =  3\nu_3 \nonumber\\ & & C_{111}  =  \nu_2,
C_{211}^{(a)}  = 2\nu_3, C_{211}^{(b)} = 2\nu_2^2, C^{(a)}_{311} =
 3\nu_4, C^{(b)}_{311} = 6\nu_3 \nu_2, C^{(c)}_{311} = 12\nu_3 \nu_2 ,
 C^{(d)}_{311} = 6\nu_2^3,  \nonumber\\
& & C^{(a)}_{411}  =  4 \nu_5, C^{(b)}_{411} = 24 \nu_3 \nu_2^2,
C^{(c)}_{411} = 24 \nu_3^2,  C^{(d)}_{411} = 24\nu_4\nu_2,
C^{(e)}_{411} = 24\nu_2^4, C^{(f)}_{411} = 96\nu_3\nu_2^2,
\nonumber\\ & & C^{(g)}_{411} = 24\nu_4\nu_2, C^{(h)}_{411} =
12\nu_3^2, C^{(i)}_{411} = 24\nu_3\nu_2^2 \nonumber\\ & & C_{1111}
=  \nu_3, C^{(a)}_{2111}= 2\nu_4, C^{(b)}_{2111} = 6 \nu_2 \nu_3
 \nonumber\\ & & C^{(a)}_{3111}= 6\nu_2 \nu_4, C^{(b)}_{3111} =
18\nu_2 \nu_4, C^{(c)}_{3111} = 18\nu_3 \nu_2^2, C^{(d)}_{3111} =
18\nu_3 \nu_2^2, C^{(e)}_{3111} = 3 \nu_5, C^{(f)}_{3111} =
18\nu_3^2, {\rm etc.}
\end{eqnarray}
The superscripts $a$, $b$, etc. denote different configurations of
internal vertices with fixed external legs and the coefficients in
each term represent the number  number of possible internal
rearrangements. We denote the amplitudes associated with $n^{th}$
order vertex of matter correlation function by $\nu_n$.

The sum of all internal vertices appearing in the
tree-representation of MCCs $\langle \delta({\bf x}_a)^p \dots
\delta({\bf x}_h)^w \rangle$ will satisfy the following equation,
which guarantees that only the leading-order diagrams with the
minimum numbers of external legs are considered:
\begin{equation}
\sum_i^p i + \sum_j^q j + \dots = 2(p+q + \dots) -2
\end{equation}
This means that, for one point cumulants, $\sum_i^p i = 2p -2$
and for 2CCs $\sum_i^p i + \sum_j^q j = 2(p+q) - 2$. These
are the same as the conditions met by the leading-order tree-level
contributions of cumulants and cumulant correlators (Bernardeau
1995). This can also be understood by noticing, for example, that
the expression for the  3CC contains factors $\bar \xi_a^{p-1}\bar
\xi_b^{q-1}\bar \xi_c^{r-1}$ for three individual cells and two
correlation functions connecting three different cells i.e.
$\xi_{ab}\xi_{bc}$ (or any cyclic permutation), so finally we get
the order to be equal to $2(p+q+r-3)+4 = 2(p+q+r) -2$.

\begin{figure}
\protect\centerline{
\epsfysize = 5.5truein
\epsfbox[-10 -10 653 574]
{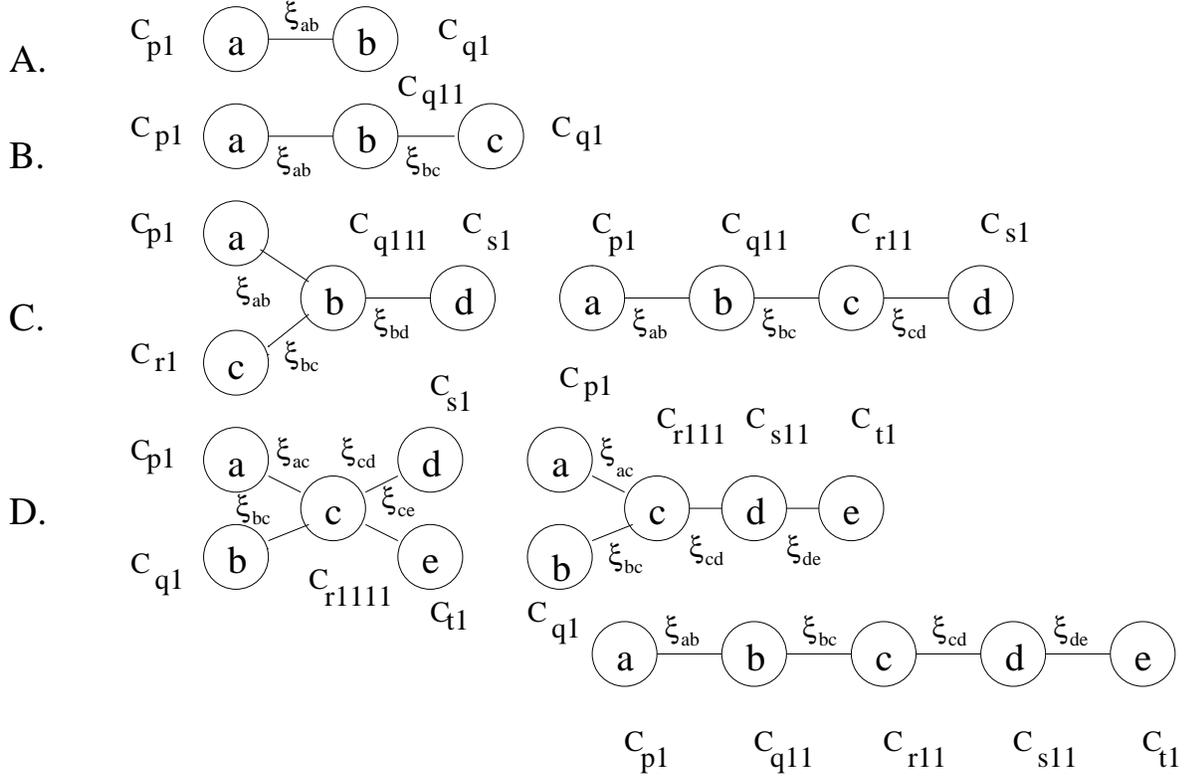} } \caption{Graphical representation of MCC
$\langle \delta(x_a)^{p} \dots \delta(x_h)^{w} \rangle$ in terms
of RCCs $C_{p1}, C_{p11}, C_{p111} \dots$. It is interesting to
note that the underlying tree structure reproduces a similar tree
hierarchy for the multi-point cumulant correlators. It is also
important to note that such a decomposition is possible if and
only if the hierarchical amplitude $\nu_n$ becomes independent of
geometrical form factors, which is true only in the highly
non-linear regime. The contribution to tree hierarchy can be
represented by different tree diagrams (Fry 1984), and all
possible topologies are considered with fixed number of vertices.
While at second order only one type of topology is possible,
higher-order contributions generate different topologies. For
example, in third order (C) one can have either a ``snake'' or a
``star'' topology.  }
\end{figure}

\begin{figure}
\protect\centerline{
\epsfysize = 5.25truein
\epsfbox[0 0 712 538]
{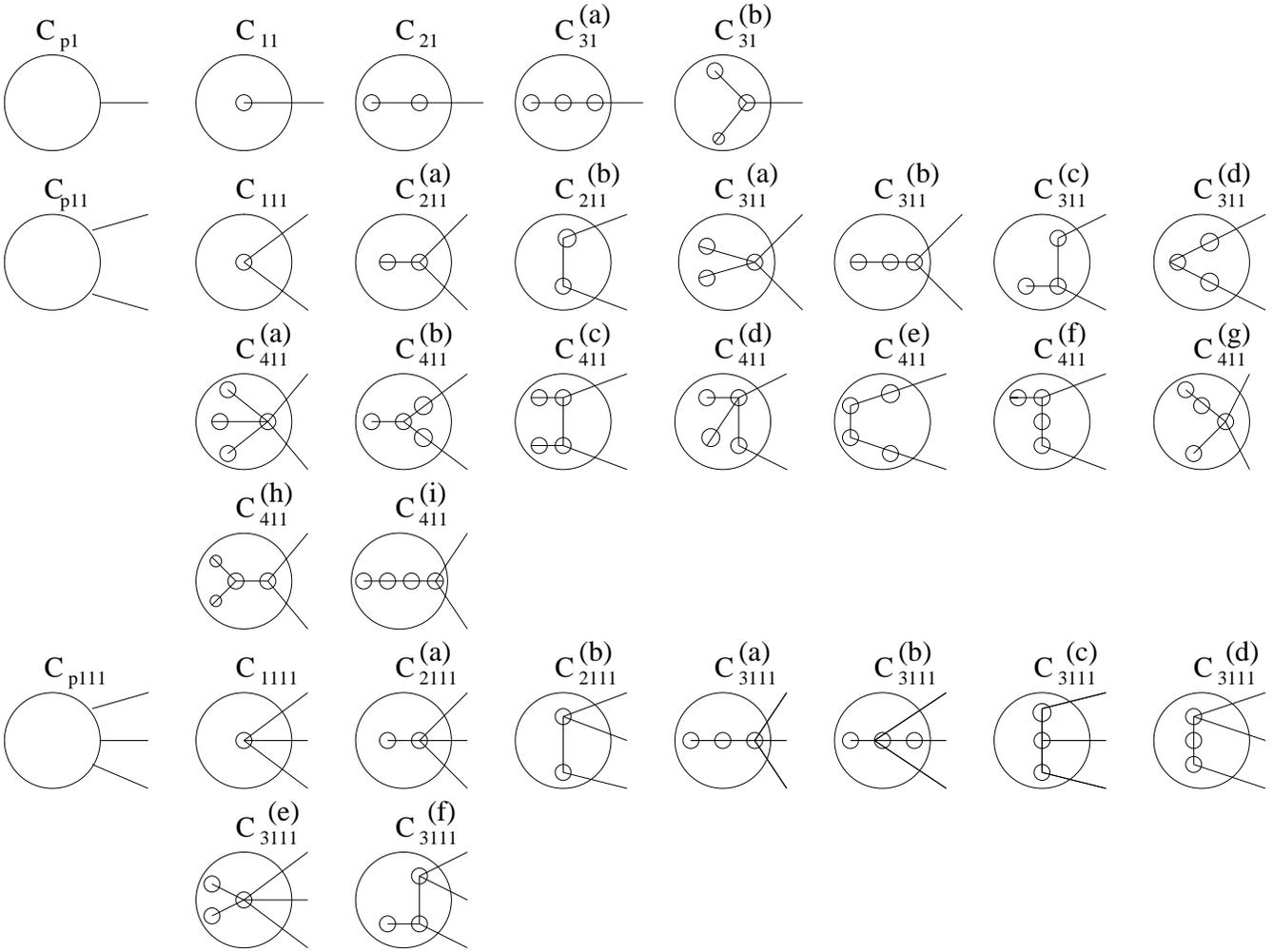} } \caption{The tree representation for
multi-point cumulants can be understood by considering the
external tree which represent how different cells are linked with
each other and the internal trees in each cells, i.e. how internal
vertices are arranged in a tree structure with fixed number of
external legs. In this figure we represent different internal
trees which appear in the multi-point cumulant correlators. }
\end{figure}

For reasons we shall discuss in the next section, we now define
generating function of the RCCs by following equations:
\begin{eqnarray}
\mu_1(y) & = & \beta(y) = \sum_{n = 1}^{\infty} {C_{n1} \over n!}
y^n = y - 2\nu_2 {y^2 \over 2!} + ( 6 \nu_2^2 + 3 \nu_3) { y^3
\over 3!} - \dots  \nonumber\\ \mu_2(y) & = & \sum_{n =
1}^{\infty} {C_{n11} \over n!}  y^n = -\nu_2 y + ( 2\nu_2^2 +
2\nu_3) {y^2 \over 2!} - ( 6 \nu_2^3 + 18 \nu_2 \nu_3 + 3 \nu_4) {
y^3 \over 3!}  \nonumber\\
 & & + ( 4\nu_5 + 48 \nu_2 \nu_4 +
 36 \nu_3^2 + 144 \nu_2^2 \nu_3 + 24 \nu_2^4) { y^4 \over 4!} +\dots
 \nonumber\\
\mu_3(y) & = & \sum_{n = 1}^{\infty} {C_{n111} \over n!} y^n =
\nu_3 y - (6\nu_2\nu_3 + 2\nu_4) {y^2 \over 2!} + ( 36
\nu_2^2\nu_3 + 18\nu_3^2 + 24\nu_2 \nu_4 + 3\nu_5) { y^3 \over 3!}
- \dots
\end{eqnarray}
The total number of trees appearing in the representation of the
$S_p$ parameters (related to $C_p$) is $p^{p-2}$. Similarly, for
$C_{p1}$, the total number of terms is $p^{p-1}$; for $C_{p11}$ it
is $p^p$, for $C_{p111}$ it is $p^{p+1}$ and $C_{p1\dots1}$ in
general will contain $p^{p+r-2}$ terms, where $r$ is the number of
external lines joining different cells in the tree representation
of the RCCs. For one point cumulants, or $S_N$ parameters, $r=0$,
for 2CCs $r=1$,  and so on.

\section{Tree representation of Multi-point Cumulants}

Let us denote the joint probability distribution function of $q$
different cells having cell occupation of $N_1, \dots, N_q$ by
$P(N_1, \dots N_q)$. The generating function of this distribution
is   ${\cal P}(\lambda_1, \dots, \lambda_q)=\exp[\chi(\lambda_1,
\dots,\lambda_q)]$.
\begin{equation}
\exp[\chi(\lambda_1, \dots,\lambda_q)] = {\cal P}(\lambda_1,
\dots, \lambda_q) = \sum_{N_1, \dots, N_q} P(N_1, \dots,N_q)
\lambda_1^{N_1} \dots \lambda_q^{N_q}.
\end{equation}
Extending the known relation for one-point statistics (Balian \&
Schaeffer 1989) and  employing the vertex generating function
\begin{equation}
G(\tau) = \sum_{n = 1}^{\infty} {\nu_n \over n!} ( - \tau)^n,
\end{equation}
Bernardeau \& Schaeffer (1992) showed that it is possible to
relate $\exp(\chi(\lambda_1, \dots,\lambda_q))$ to $G(\tau)$ using
\begin{eqnarray}
\chi(\lambda_1, \dots,\lambda_q) & = & \sum_{i=1}^q -{1 \over
\xi_i} (y_i G(\tau_i) - {1 \over 2} y_i \tau_i G'(\tau_i)) \\
\tau_i & = & -y_i G'(\tau_i) - \sum_{i\ne j} { \xi_{ij} \over \bar
\xi_j} G'(\tau_j) \label{pain}
 \end{eqnarray}
The correlation of different cells enters through $\xi_{ij}/ \bar
\xi_i$ in eq. (\ref{pain}). If we ignore this term in the second
equation, which governs the  behaviour of $\tau_i$, then the
statistics of different cells are independent of each other. This
is a kind of zeroth approximation when only one--point statistics
are used in the description. If we expand $(\chi(\lambda_1,
\dots,\lambda_q))$
 in a series of powers of $\xi_{ij}/ \bar \xi_i$, then we  will recover
the effect of correlations between different cells of
progressively higher order. Expanding eq. (\ref{pain}) we can
write:
\begin{equation}
\tau_i^{(0)} + \tau_i^{(1)} + \tau_i^{(2)} \dots + y_i
G'(\tau_i^{(0)} +  \tau_i^{(1)} + \tau_i^{(2)}) = - \sum_{i\ne j}{
\xi_{ij} \over {\bar \xi_i}} y_j G'(\tau_j^{(0)} + \tau_j^{(1)} +
\tau_j^{(2)}).
\end{equation}

Corrections up to second order in $(\xi_{ij}/\bar \xi_i)$ were
computed by Bernardeau \& Schaeffer (1992). These and the next
higher-order terms are as follows:
\begin{eqnarray}
\tau_i^{(0)} & = &  -y_i G'(\tau_i^{(0)})\nonumber \\ \tau_i^{(1)}
& = & -{1 \over [1 + y_iG''(\tau_i^{(0)})]}\sum_{i\ne j}{ \xi_{ij}
\over {\bar \xi_i}} y_jG'(\tau_j^{(0)}) = {1 \over [1 +
y_iG''(\tau_i^{(0)})]}\sum_{i\ne j}{ \xi_{ij} \over {\bar \xi_i}}
\tau_j^{(0)}\nonumber \\ \tau_i^{(2)} & = & -{1 \over [1 +
y_iG''(\tau_i^{(0)})]} \Big [ y_i {{\tau_i^{(1)}}^2 \over 2!}
G'''(\tau_i^{(0)})  + \sum_{i\ne j}{\xi_{ij} \over {\bar \xi_i}}
y_j G''(\tau_j^{(0)}) \tau_j^{(1)} \Big ] \nonumber \\
\tau_i^{(3)} & = & -{1 \over [1 + y_iG''(\tau_i^{(0)})]} \Big [
y_i {\tau_i^{(1)} {\tau_i^{(2)}}} G'''(\tau_i^{(0)})  + y_i
{{\tau_i^{(1)}}^3 \over 3!}G^{IV}(\tau_i^{(0)}) - \sum_{i\ne
j}{\xi_{ij} \over {\bar \xi_i}} y_j G''(\tau_j^{(0)}) \tau_j^{(2)}
+ \sum_{i\ne j}{\xi_{ij} \over {\bar \xi_i}} y_j
G'''{(\tau_j^{(0)})}^2 {\tau_j^{(1)}}^2 \Big ]\nonumber
\\ \tau_i^{(4)} & = & -{1 \over [1 + y_iG''(\tau_i^{(0)})]} \Big [
{{\tau_i^{(2)}}^2 \over 2!} G'''(\tau_i^{(0)}) +
{{\tau_i^{(1)}}^2{\tau_i^{(2)}} \over 2!} G^{IV}(\tau_i^{(0)}) +
{{\tau_i^{(1)}}^4 \over 4!} G^{V}(\tau_i^{(0)}) \nonumber\\ &
&+\sum_{i\ne j}{\xi_{ij} \over {\bar \xi_i}} y_j \Big [
G''(\tau_j^{(0)}) \tau_j^{(3)} + G'''(\tau_j^{(0)}) \tau_j^{(1)}
\tau_j^{(2)} + G^{IV}(\tau_j^{(0)}) {{\tau_j^{(1)}}^3 \over 3!} )
\Big ].
\end{eqnarray}
Certain predictions of such a series expansion have already been
tested against numerical simulations, at least to linear order.
Such predictions from the tree-level hierarchy relate to the
two--point cumulant correlators (2CCs) which satisfy the relation
$\langle \delta^p({\bf x}_1) \delta^q({\bf x}_2 \rangle = [C_{p1}
C_{q1}\xi_{ab}]{\bar \xi}^{p+q -2}$ (Bernardeau \& Schaeffer 1992;
Bernardeau 1995). This property of factorization is  related to
the fact that the bias associated with an overdense object is an
intrinsic property of the object. The bias associated with a pair
comprising two different objects can be factorised into two
separate quantities which each represent the intrinsic bias of one
constituent. In recent work Munshi \& Melott (1998) were able to
show that predictions such as $C_{pq}  = C_{p1}C_{q1}$ are
satisfied very accurately in highly non-linear regime when
two over dense cells are not separated by a large distance and
$(\xi_{ij}/ \bar \xi_i) < 1$ is satisfied. Other related
predictions regarding bias have also been tested successfully by
Munshi et al. (1998). The Hierarchical ansatz implies  natural
predictions for other higher order quantities, such as one point
cumulants for collapsed objects which we will also compute here. A
comparison of such predictions against numerical simulations is
left for future work.

In a similar vein, we can consider a series expansion of
$\chi(\lambda_1, \dots, \lambda_q)$ in $(\xi_{ij}/ \bar \xi_i)$ we
get higher order correction terms that involve contribution from
the $\tau_i$ to the same order. In leading order it can be shown
that all over-dense cells are uncorrelated. Each higher-order term
encodes the effect of one more extra neighbouring cells on
occupancy of a given single cell.
\begin{eqnarray}
\chi^{(0)}(\lambda_i) & = & -\sum {y_i \over \bar \xi_i}  G(\tau_i^{(0)}) +
{1 \over 2 } \sum { y_i \over \bar \xi_i} \tau_i^{(0)}
G'({\tau_i^{(0)}})\nonumber \\ \chi^{(1)}(\lambda_i) & = &  -\sum {y_i \over
\bar \xi_i}  G'(\tau_i^{(0)}) \tau_i^{(1)} + {1 \over 2}\sum {y_i
\over \bar \xi_i}  G'(\tau_i^{(0)}) \tau_i^{(1)} +  {1 \over
2}\sum {y_i \over \bar \xi_i}  G''(\tau_i^{(0)}) \tau_i^{(1)}
\tau_i^{(0)}\nonumber \\ \chi^{(2)}(\lambda_i) & =&   -\sum {y_i \over \bar
\xi_i}  G'(\tau_i^{(0)}) \tau_i^{(2)} + {1 \over 2}\sum {y_i \over
\bar \xi_i}  G'(\tau_i^{(0)}) \tau_i^{(2)} + {1 \over 2}\sum {y_i
\over \bar \xi_i}  {G'''(\tau_i^{(0)})} {\tau_i^{(0)} \over 2!}
{\tau_i^{(1)}}^2+ {1 \over 2}\sum {y_i \over \bar \xi_i}
G''(\tau_i^{(0)}) \tau_i^{(2)}\tau_i^{(0)} \nonumber \\
\chi^{(3)}(\lambda_i)
& = &  -\sum {y_i \over \bar \xi_i} G'(\tau_i^{(0)}) \tau_i^{(3)}
- \sum {y_i \over \bar \xi_i} {G'''(\tau_i^{(0)})}
{{\tau_i^{(1)}}^3 \over 3!}
 + {1 \over 2}\sum {y_i \over \bar \xi_i}  {G^{IV}(\tau_i^{(0)})}\tau_i^{(0)}
 {{\tau_i^{(1)}}^3 \over 3!}
+ {1 \over 2}\sum {y_i \over \bar \xi_i} \tau_i^{(0)}
{G'''(\tau_i^{(0)})} {\tau_i^{(1)}}{\tau_i^{(2)} }\nonumber\\ & &+
~{1 \over 2}\sum {y_i \over \bar \xi_i}  {G'''(\tau_i^{(0)}) }
{{\tau_i^{(1)}}^3 \over 2!} + {1 \over 2}\sum {y_i \over \bar
\xi_i}  {G''(\tau_i^{(0)}}) {\tau_i^{(0)}}  {\tau_i^{(3)}} + {1
\over 2}\sum {y_i \over \bar \xi_i}  {G'(\tau_i^{(0)}})  {\tau_
i^{(3)}}\nonumber \\\nonumber \chi^{(4)}(\lambda_i) & = & -\sum {y_i \over
\bar \xi_i} G'(\tau_i^{(0)}) \tau_i^{(4)} -\sum {y_i \over \bar
\xi_i}  G^{IV}(\tau_i^{(0)}) {{\tau_i^{(1)}}^4 \over 4!}
\nonumber\\& & + {1 \over 2} \sum {y_i \over \bar \xi_i}
G^{V}(\tau_i^{(0)}) {{\tau_i^{(1)}}^4 \over 4!} + {1 \over 2} \sum
{y_i \over \bar \xi_i}  G^{IV}(\tau_i^{(0)}) {{\tau_i^{(1)}}^2
\tau_i^{(2)}}{ \tau_i^{(0)} \over 2!} + {1 \over 2} \sum {y_i
\over \bar \xi_i}
 G^{IV}(\tau_i^{(0)}) {{\tau_i^{(1)}}^3}{ \tau_i^{(1)} \over 3!} \nonumber
 \\
& & + {1 \over 2} \sum {y_i \over \bar \xi_i}  G'''(\tau_i^{(0)})
{{\tau_i^{(1)}}^2}{ \tau_i^{(2)} \over 2!}
 + {1 \over 2} \sum {y_i \over \bar \xi_i}  G'''(\tau_i^{(0)}) {{\tau_i^{(2)}}^2}
 { \tau_i^{(0)} \over 2!} + {1 \over 2} \sum {y_i \over \bar \xi_i}
 G''(\tau_i^{(0)}) {{\tau_i^{(4)}}}{ \tau_i^{(0)}} \nonumber \\
&&  + {1 \over 2}
 \sum {y_i \over \bar \xi_i}  G'(\tau_i^{(0)}) {{\tau_i^{(4)}}}
\end{eqnarray}
After a tedious but straightforward exercise in algebra it is
possible to express every order of $\chi(\lambda_1 \dots
\lambda_q)$ in terms of $\tau_i^{(0)}$. It is interesting to note
that every new power of $(\xi_{ij}/\bar \xi_i)$ in the expansion
of $\chi(\lambda_1 \dots \lambda_q)$ adds a new ``star'' vertex of
same order. What is most remarkable however is that such an
expansion is also able to reproduce other diagrams such as the
``snake'' and hybrid of ``stars'' and ``snakes'' of lower orders.
\begin{eqnarray}
\chi^{(0)}( \lambda_i)  & = &  -\sum_{i = 1}^q { y_i \over \xi_i}
G(\tau_i^{(0)}) + {1 \over 2} \sum {y_i \over \xi_i} \tau_i^{(0)}
G'(\tau_i^{(0)})\nonumber \\ \chi^{(1)}( \lambda_i) & = & {1 \over
2} \sum_i \sum_{i \ne j} { \tau_i^{(0)} \over \bar \xi_i} \xi_{ij}
{ \tau_j^{(0)} \over \bar \xi_j}\nonumber \\ \chi^{(2)}(
\lambda_i) & = & {1 \over 2} \sum_i \sum_{i \ne j} \sum_{i \ne k
}{ \tau_i^{(0)} \over \bar \xi_i} \xi_{ij} { \mu_2(\tau_j^{(0)})
\over \bar \xi_j} \xi_{jk} { \tau_k^{(0)} \over \bar
\xi_k}\nonumber \\ \chi^{(3)}( \lambda_i) & = & {1 \over 3!}
\sum_i \sum_{i \ne j} \sum_{i \ne k} \sum_{i \ne l} {\mu_3
(\tau_i^{(0)}) \over \bar \xi_i }{\xi_{ij}} {\tau_j^{(0)} \over
\bar \xi_j } {\xi_{ik}} {\tau_k^{(0)} \over \bar \xi_k }{\xi_{il}}
{\tau_l^{(0)} \over \bar \xi_l }~~~~~~~~~~~~~~~~~~~~~~~~~{\rm
Star~ Topology~(4)}\nonumber\\ & &+ {1 \over 2} \sum_i \sum_{i \ne
j} \sum_{j \ne k} \sum_{k \ne l} {\tau_i^{(0)} \over \bar \xi_i }
\xi_{ij} {\mu_2(\tau_j^{(0)}) \over \bar \xi_j }
\xi_{jk}{\mu_2(\tau_k^{(0)}) \over \bar \xi_j }
\xi_{kl}{\tau_l^{(0)} \over \bar \xi_l}~~~~~~~~~~~~~~~~{\rm Snake~
Topology~(12)} \nonumber \\
 \chi^{(4)} ( \lambda_i) & = &  {1 \over 4!} \sum_i \sum_{i \ne j} \sum_{i \ne k} \sum_{i \ne l} {\mu_i^{(4)}(\tau_i^{(0)}) \over \bar \xi_i } {\xi_{ij}} {\tau_j^{(0)} \over \bar \xi_j }
{\xi_{ik}} {\tau_k^{(0)} \over \bar \xi_k }{\xi_{il}}
{\tau_l^{(0)} \over \bar \xi_l } {\xi_{im}} {\tau_m^{(0)} \over
\bar \xi_m }~~~~~~~~~~~~~~{\rm Star~ Topology~(5)}\nonumber \\ &&
+ {1 \over 2!} \sum_i \sum_{i \ne j} \sum_{j \ne k} \sum_{k \ne l}
{\tau_i^{(0)} \over \bar \xi_i} {\xi_{ij} } {\mu^{(2)}(
\tau_j^{(0)}) \over \bar \xi_j } {\xi_{jk} }
{\mu^{(2)}(\tau_k^{(0)}) \over \bar\xi_k } {\xi_{kl}} {\mu^{(2)}(
\tau_l^{(0)}) \over \bar \xi_l} {\xi_{lm}} { \tau_m^{(0)} \over
\bar \xi_m} ~~~~~~{\rm Hybrid~ Topology~(60)}\nonumber \\ & & + {1
\over 2!} \sum_i \sum_{i \ne j} \sum_{j \ne k} \sum_{k \ne l}
{\tau_i^{(0)} \over \bar \xi_i} {\xi_{ij} } {\mu^{(2)}(
\tau_j^{(0)}) \over \bar \xi_j } {\xi_{jk} }
{\mu^{(2)}(\tau_k^{(0)}) \over \bar\xi_k } {\xi_{kl}} {\mu^{(2)}(
\tau_l^{(0)}) \over \bar \xi_l} {\xi_{lm}} { \tau_m^{(0)} \over
\bar \xi_m}~~~~~~{\rm Snake~ Topology~(60)}.
\end{eqnarray}
We have reintroduced quantities $\mu_n$ that represent vertices in
the tree-level diagrams of MCCs, $n$ representing the number of
external legs associated with such vertices. In terms involving
ordered sums we can replace $\displaystyle{ 1 \over 2} \sum_i
\sum_{i \ne j}$ by $\displaystyle ,\sum_{\rm pairs}$ (i.e. sum
over all pairs) and similarly $\displaystyle {3 \over 3!} \sum_i
\sum_{i \ne j} \sum_{i \ne k}$ by $\displaystyle \sum_{\rm
triplets}$ (i.e. sum over all triplets) and additional terms which
contribute to loop corrections to lower order terms. Similarly
$\displaystyle {4 \over 4!} \sum_i \sum_{i \ne j} \sum_{i \ne
k}\sum_{i \ne l}$ can be replaced by $\displaystyle \sum_{\rm
quadruplets}$ and loop contribution to lower order terms.

It is also pertinent to note that, at every order, the
hierarchical ansatz reproduces the underlying tree structure of
matter correlation, but with different amplitudes for the
vertices. It also produces different loop corrections to lower
order diagrams. Since we are only interested in the leading order
contributions to the statistics of collapsed objects (important in
large separation limit) we will only focus on tree level diagrams.
If we however focus on statistics of the initial Gaussian field it
is important to note that since the reduced correlation functions
of odd orders are zero,  one needs to consider these corrective
terms which are neglected here as in this case they will provide
the  dominant contributions.

The expression of $\mu_n$ to arbitrary order can be used to
construct statistics of collapsed objects, such as the probability
distribution function or the void probability function, quantities
which are generally easier to extract from numerical simulations
than the higher order moments themselves. We begin with:
\begin{eqnarray}
\mu_1(y)  & = &  -y G'(\tau(y)) \nonumber\\ \mu_2(y)  & = & { -y
G''(\tau(y)) \over 1 + y G''(\tau(y))} \nonumber \\ \mu_3(y) & = &
{-y G'''(\tau(y)) \over ( 1 + y G''(\tau(y))^3} \nonumber \\
\mu_4(y) & = & {-y G^{IV}(\tau(y)) \over [ 1 + y G''(\tau(y))]^4}
+ {3 y^2 G'''(\tau(y))^2 \over [ 1 + y G''(\tau(y)]^5} \nonumber \\
\mu_5(y) & = & {-y G^{V}(\tau(y)) \over [ 1 + y G''(\tau(y))]^5 }
+ {10 y^2 G^{IV}(\tau(y)G'''(\tau(y)) \over [ 1 + y
G''(\tau(y))]^6} - {15 y^3 G'''(\tau(y))^3 \over [ 1 + y G'' (\tau(y))^7}.
\end{eqnarray}
The series expansion of these functions will reproduce terms which
we have earlier obtained from counting internal degrees of
freedoms associated with different nodes in tree representation of
the RCCs. One can then construct the successive approximations to
${\cal P}$ as follows:
\begin{eqnarray}
{\cal P}^{(0)}(\lambda_1, \dots, \lambda_q) & = & \prod_i {\cal
P}(\lambda_i)\nonumber \\ {\cal P}^{(1)} (\lambda_1, \dots,
\lambda_q) & = & \sum_{(i,j)}{ \tau_i^{(0)} \over \bar \xi_i}{\cal
P}(\lambda_i)\xi_{ij}{\cal P}(\lambda_j) { \tau_j^{(0)} \over \bar
\xi_j} \prod_{k \ne i,j} {\cal P}(\lambda_k)\nonumber \\ {\cal
P}^{(2)} (\lambda_1, \dots, \lambda_q)& = & \sum_{(i,j,k)}{
\tau_i^{(0)} \over \bar \xi_i}{\cal P}(\lambda_i)\xi_{ij} {
\mu_2(y) \over \bar \xi_j}{\cal P}(\lambda_j)  \xi_{jk}{
\tau_k^{(0)} \over \bar \xi_k} {\cal P}(\lambda_ k) \prod_{l \ne
i,j,k} {\cal P}(\lambda_l)\nonumber \\ {\cal P}^{(3)} (\lambda_1,
\dots, \lambda_q) & = & \sum_{(i,j,k,l)}{ \mu_3(y) \over \bar
\xi_i}{\cal P}(\lambda_i)\xi_{ij} { \tau_j^{(0)} \over \bar
\xi_j}{\cal P}(\lambda_j)  \xi_{ik}{ \tau_k^{(0)} \over \bar
\xi_k}{\cal P}(\lambda_k) \xi_{il}{ \tau_l^{(0)} \over \bar
\xi_l}{\cal P}(\lambda_l)\prod_{m \ne i,j,k,l}{\cal P}(\lambda_m)
 \nonumber\\ && + \sum_{(i,j,k,l)}{ \tau_i^{(0)} \over \bar \xi_i}
{\cal P}(\lambda_i)\xi_{ij} {\mu_2(y) \over \bar
\xi_j}P(\lambda_j)  \xi_{jk}{\mu_2(y) \over \bar \xi_k}{\cal
P}(\lambda_k) \xi_{il}{ \tau_l^{(0)} \over \bar \xi_l}{\cal
P}(\lambda_l)\prod_{m \ne i,j,k,l}{\cal P}(\lambda_m) \nonumber \\
{\cal P}^{(4)} (\lambda_1, \dots, \lambda_q) & = &
\sum_{(i,j,k,l,m)}{ \mu_4(y) \over \bar \xi_i}{\cal
P}(\lambda_i)\xi_{ij} { \tau_j^{(0)} \over \bar \xi_j}{\cal
P}(\lambda_j)  \xi_{ik}{ \tau_k^{(0)} \over \bar \xi_k}{\cal
P}(\lambda_k) \xi_{il}{ \tau_l^{(0)} \over \bar \xi_l}{\cal
P}(\lambda_l)\xi_{im}{ \tau_m^{(0)} \over \bar \xi_m}{\cal
P}(\lambda_m)\prod_{r \ne i,j,k,l,m} {\cal P}(\lambda_r)
\nonumber\\ & & + \sum_{(i,j,k,l,m)}{ \tau_i^{(0)} \over \bar
\xi_i} {\cal P}(\lambda_i)\xi_{ij} {\mu_3(y) \over \bar
\xi_j}{\cal P}(\lambda_j) \xi_{jk}{\tau_k^{(0)} \over \bar
\xi_k}{\cal P}(\lambda_k) \xi_{jl}{ \mu_2(y) \over \bar
\xi_l}{\cal P}(\lambda_l)\xi_{lm} { \tau_m^{(0)} \over \bar
\xi_m}{\cal P}(\lambda_m) \prod_{r \ne i,j,k,l,m}{\cal
P}(\lambda_r)\nonumber \\ && + \sum_{(i,j,k,l,m)}{ \tau_i^{(0)}
\over \bar \xi_i} {\cal P}(\lambda_i)\xi_{ij} {\mu_2(y) \over \bar
\xi_j}P(\lambda_j)  \xi_{jk}{\mu_2(y) \over \bar \xi_k}{\cal
P}(\lambda_k) \xi_{kl}{ \mu_2(y) \over \bar \xi_l}{\cal
P}(\lambda_l)\xi_{lm}{ \tau_m^{(0)} \over \bar \xi_m}{\cal
P}(\lambda_m) \prod_{r \ne i,j,k,l,m} {\cal P}(\lambda_r)
\end{eqnarray}

The quantity ${\cal P}(\lambda)$ plays the role of the generating
function for the one-point probability distribution $P(N)$. If one
incorporates the ``bias'' for a cell of occupation number $N$ by
$b(N)$ then the generating function for $P(N)b(N)$ can be seen to
be ${\cal P}(\lambda) \tau(y) / \bar \xi$. Similarly, ${\cal P}
(\lambda) \mu_s(y) / \bar \xi$ is the generating function for
$P(N)\nu_s(N)$. To summarise:
\begin{eqnarray}
\sum \lambda^N P(N) & = & {\cal P}(\lambda);\nonumber\\ \sum
\lambda^N P(N)b(N) & = & {\cal P}(\lambda) {\tau(\lambda) \over
\bar \xi_2};\nonumber\\  \sum \lambda^N P(N) \nu_2(N) & = & {\cal
P}(\lambda) {\mu(\lambda) \over \bar \xi_2};  \nonumber\\ \sum
\lambda^N P(N) \nu_s(N) & = & {\cal P}(\lambda) {\mu_s(\lambda)
\over \bar \xi_2}.
\end{eqnarray}
Using the generating functions we have obtained (21), we can then
express the count probability distribution function to
successive order:
\begin{eqnarray}
P^{(0)}({\bf N}) & = & \prod_{i} P(N_i); \nonumber\\ P^{(1)}({\bf
N}) & = & \sum_{k \ne i,j} P(N_k) \prod_{(i,j \ne
k)}b(N_j)P(N_j)\xi_{ij}P(N_i)b(N_i);\nonumber \\ P^{(2)}({\bf N})
& = & \sum_{l \ne i,j,k} P(N_l) \prod_{(i,j,k
\ne
l)}\nu_2(N_i)P(N_i)\xi_{ij}P(N_j)b(N_j)\xi_{jk}P(N_k)b(N_k);\nonumber
\\ P^{(3)}({\bf N}) & = & \sum_{m \ne i,j,k,l} P(N_m)
\prod_{(i,j,k,l \ne
m)}\nu_3(N_i)P(N_i)\xi_{ij}P(N_j)b(N_j)\xi_{ik}P(N_k)b(N_k)
\xi_{il}P(N_l)b(N_l)  \nonumber\\ & &+ \sum_{r \ne i,j,k,l,m}
P(N_m) \prod_{(i,j,k,l \ne
m)}P(N_i)b(N_i)\xi_{ij}\nu_2(N_j)P(N_j)\xi_{jk}\nu_2(N_k)P(N_k)
\xi_{kl}P(N_l)b(N_l);\nonumber\\ P^{(4)}({\bf N}) & = & \sum_{r
\ne i,j,k,l,m} P(N_r) \prod_{(i,j,k,l,m \ne
r)}\nu_4(N_i)P(N_i)\xi_{ij}P(N_j)b(N_j)\xi_{ik}P(N_k)b(N_k)
\xi_{il}P(N_l)b(N_l) \xi_{im}P(N_m)b(N_m)\nonumber\\  &&+ \sum_{r
\ne i,j,k,l,m} P(N_m) \prod_{(i,j,k,l \ne
m)}P(N_i)b(N_i)\xi_{ij}\nu_3(N_j)P(N_j)\xi_{jk}P(N_k)b(N_k)
\xi_{jl}\nu_2(N_l)P(N_l)\xi_{lm}P(N_m)b(N_m)\nonumber \\ &&+
\sum_{r \ne i,j,k,l,m} P(N_m) \prod_{(i,j,k,l \ne
m)}P(N_i)b(N_i)\xi_{ij}\nu_2(N_j)P(N_j)\xi_{jk}\nu_2(N_k)P(N_k)
\xi_{kl}\nu_2(N_l)P(N_l)\xi_{lm}P(N_m)b(N_m).
\end{eqnarray}
where we have used the notation $P^{(i)}({\bf
N})=P^{(i)}(N_1,\ldots, N_q)$ for arbitrary number of cells $q$.
The product terms involve pairs for $i=1$, triplets for $i=2$,
quadruplets for $i=3$ and pentuplets for $i=4$. The  functions
$b(N) = \mu_1(N)$ and $\mu_N(N)$, which describe the effect of
correlations between different cells, will each satisfy two
normalization conditions  satisfied by the multi-point count
probability distribution function (CPDF) $P(N_1, \dots, N_k)$. The
first set of these conditions is obtained by taking the first
order moment of $P(N_1, \dots, N_k)$ with respect to all but one
of the $N_k$, and the second set is obtained by taking moments
with respect to all $N_k$ and using the definition of the
underlying matter correlation function. Hence we impose
\begin{equation}
\sum P(N)\nu_s(N) = 0; ~~~~~~~~~\sum NP(N)\nu_s(N) = \bar N \nu_s,
\end{equation}
where $\nu_N$ are the amplitudes associated with vertices of
underlying mass distribution in the highly nonlinear regime.

Using inverse Laplace transforms, it is now possible to relate the
distributions $P(N)$ or $P(N)\nu_s(N)$ in terms of their
generating functions $P(\lambda)$ or in terms of $\mu_s(y)$. In
turns out that the results can be expressed in terms of functions
of a scaling variable $x = N/N_c$, where $N_c = \bar N \bar \xi_2$
and $\bar N$ is the typical occupation number of cells under
consideration.
\begin{eqnarray}
P(N) & = & { 1 \over 2 \pi i} \int { d \lambda  \over
\lambda^{N+1}} {\cal P}(\lambda) = { 1 \over \xi_2 N_c}
h(x)\nonumber
\\ P(N)b(N) & = & { 1 \over 2 \pi i} \int { d \lambda  \over
\lambda^{N+1}} {\cal P}(\lambda) {\tau(y) \over \bar \xi} = { 1
\over \xi_2 N_c} h(x)b(x)\nonumber \\ P(N)\nu_2(N) & = & { 1 \over
2 \pi i} \int { d \lambda \over \lambda^{N+1}} {\cal P}(\lambda)
{\mu_2(y) \over \bar \xi} = { 1 \over \xi_2 N_c}
h(x)\mu_2(x)\nonumber
\\ P(N)\nu_3(N) & = & { 1 \over 2 \pi i} \int { d \lambda  \over
\lambda^{N+1}} {\cal P}(\lambda) {\mu_3(y) \over \bar \xi} = { 1
\over \xi_2 N_c} h(x)\mu_3(x).
\end{eqnarray}

In turn the scaling functions $h(x)$ and $b(x)$, which describe
the behaviour of $P(N)$, $b(N)$ and $\nu_s(N)$ for different
length scales and for different levels of non-linearity, are
expressed in terms of the vertex-generating functions $\mu_s(y)$
we introduced earlier in equation (21):

\begin{eqnarray}
h(x) & = & -{1 \over 2\pi i}\int_{i\infty}^{i\infty} dy~y\sigma(y)
\exp(yx);\nonumber\\ h(>x) & = & {1 \over 2\pi
i}\int_{i\infty}^{i\infty} dy~\sigma(y) \exp(yx); \nonumber \\
\nu_1(x)h(x) & = & b(x)h(x) = -{1 \over 2\pi
i}\int_{i\infty}^{i\infty} dy~\tau(y) \exp(yx); \nonumber\\
\nu_1(>x)h(>x) &  = & b(>x)h(>x) = {1 \over 2\pi
i}\int_{i\infty}^{i\infty} dy {\tau(y) \over y} \exp(yx);
\nonumber
\\ \nu_2(x)h(x) & = & -{1 \over 2\pi i}\int_{i\infty}^{i\infty}
dy~\mu_2(y) \exp(yx); \nonumber \\ \nu_2(>x)h(>x) & = & {1 \over
2\pi i}\int_{i\infty}^{i\infty} dy {\mu_2(y) \over y}
\exp(yx);\nonumber\\ \nu_3(x)h(x) & = & -{1 \over 2\pi
i}\int_{i\infty}^{i\infty} dy~\mu_3(y) \exp(yx); \nonumber
\\ \nu_3(>x)h(>x) & = & {1 \over 2\pi i}\int_{i\infty}^{i\infty} dy
{\mu_3(y) \over y} \exp(yx),
\end{eqnarray}
in which the further functions  $h(>x), \nu_s(>x)$, etc.  are
defined by
\begin{eqnarray}
h(>x) & = & \int_x^{\infty} h(x')dx';\nonumber \\ b(>x) & = &{
\int_x^{\infty} h(x')b(x')dx' \Big / \int_x^{\infty} h(x')dx'};
\nonumber\\ \nu_2(>x) & = & { \int_x^{\infty} h(x')\nu_2(x')dx'
\Big / \int_x^{\infty} h(x')dx'};\nonumber
\\ \nu_3(>x) & = & { \int_x^{\infty} h(x')\nu_3(x')dx' \Big /
\int_x^{\infty} h(x')dx'};\nonumber \\ \nu_4(>x) & = & {
\int_x^{\infty} h(x')\nu_4(x')dx' \Big / \int_x^{\infty}
h(x')dx'}.
\end{eqnarray}
One of the reasons for using the cumulative scaling functions
(such as $h(>x)$ or $\mu(>x)$) instead of $h(x)$ and $\mu_s(x)$ is
that one can more easily incorporate a threshold in cell
occupation numbers to define the overdense objects.  Cell
statistics defined in  this way are much more stable than
differential count statistics and easier to test against numerical
simulations, as they depend less on specific models for underlying
mass distribution characterized by $G(\tau)$.

\section{Multi-Point Cumulants and Statistics of Collapsed Objects}

We can now  discuss the statistics of overdense cells in terms of
their vertex representation. Using the results of the previous
section, we can assign a weight $M_s$ to vertices of the trees
representing the distribution of overdense cells, such that with
$M_s$ plays the same role as $\nu_s$ in the tree expansion for the
underlying distribution. A complete knowledge of all the nodes for
overdense cells (or their generating function)  guarantees a full
statistical description for collapsed objects, including the
one-point cumulants $S_N^c$  and cumulant correlators $C_{NM}^c$,
the subscripting referring to the collapsed objects which we
assume can be selected by applying a density threshold to the
cells . In the following calculations give expressions for orders
$\le 5$. The quantities $M_s$ can be written
\begin{equation}
M_s(N) = {\nu_s(N) \over b(>N)^2};~~~M_s(>N) = {\nu_s(N) \over
b(>N)^2}.
\end{equation}
We can write the one-point cumulants in terms of the amplitudes
$M_s$ as follows, which are similar to the case of underlying mass
distribution:
\begin{eqnarray}
S_3^c(>x) & = & 3 {\nu_2(>x) \over b(>x)^2} = 3M_2(>x) \nonumber\\
S_4^c(>x) & = & 4 {\nu_3(>x) \over b(>x)^3} + 12 \Big ( {\nu_2(>x)
\over b(>x)^2} \Big)^2\nonumber\\ & = & 4M_3(>x) + 12
M_2(>x)^2\nonumber\\ S_5^c(>x) & = & 5 {\nu_3(>x) \over b(>x)^3} +
60 {\nu_3(>x) \over b(>x)^3}\Big ( {\nu_2(>x) \over b(>x)^2}
\Big)^2 + 60 \Big ( {\nu_2(>x) \over b(>x)^2} \Big)^3\nonumber\\ &
= & 5M_4(>x) + 60 M_3(>x) M_2(>x)+ 60 M_2(>x)^3.
\end{eqnarray}
The two-point cumulants are given by
\begin{eqnarray}
C_{21}^c(>x) & = & 2 {\mu_2(>x) \over b(>x)^2} = 2M_2(>x)\nonumber
\\ C_{31}^c(>x) & = & 3 {\mu_3(>x) \over b(>x)^3} + 6 \Big (
{\mu_2(>x) \over b(>x)^2} \Big)^2\nonumber\\
 & = & 3M_3(>x) + 6
M_2(>x)^2 \nonumber\\ C_{41}^c(>x) & = & 4 {\mu_4(>x) \over
b(>x)^3} + 36 {\mu_3(>x) \over b(>x)^3}\Big ( {\mu_2(>x) \over
b(>x)^2} \Big)^2 + 24 \Big ( {\mu_2(>x) \over b(>x)^2} \Big)^3
\nonumber\\ & = & 4M_4(>x) + 36 M_3(>x) M_2(>x) + 24 M_2(>x)^3.
\end{eqnarray}
In the following two subsections we obtain approximate forms for
the amplitudes $M_s$ in two different limiting cases. For the
determination of $S_N$ and $C_{NM}$ parameters for all values of
mass it is necessary to evaluate contour integrals relating
$\nu_2(x)$ and $\mu_2(y)$ numerically, an exercise we do not
attempt here.

\subsection{The case of moderately over-dense objects ($x <<1$)}

The asymptotic forms of the different scaling functions in the
limit  $x<<1$ are depend upon the limiting behaviour of the
scaling functions when $y >> 1$. In this  case, the scaling
function $\sigma(y)$ is known to exhibit a power law profile with
amplitude $a$ and power-law index $\omega$ both depending on
initial power spectral index (for a detail derivation see Balian \&
Schaeffer 1989, and Bernardeau \& Schaeffer 1992):
\begin{equation}
\sigma(y) = a y^{-\omega};  h(x) = a{( 1 - \omega) \over
\Gamma(\omega)} x^{\omega - 2},
\end{equation}
where $\Gamma(\cdot)$ is the $\Gamma$--function, and we use the specific
form of $G(\tau)$ given by Bernardeau \& Schaeffer (1992):
\begin{equation}
 G(\tau) = { \big ( 1 + {\tau \over \kappa}\big )^{-\kappa}}.
\end{equation}
The asymptotic behaviour of $G(\tau)$ can then be expressed as
\begin{equation}
G(\tau) \propto c \tau^{-\kappa},
\end{equation}
and this can be used to compute the asymptotic form for
$\sigma(y)$:
\begin{eqnarray}
\tau(y) & = & (yc\kappa)^{1 \over (\kappa + 2)}  \nonumber\\
\sigma(y) & = & ( 1 + {\kappa \over 2} ) \kappa^{ -\kappa \over
(\kappa + 2)} c^{1-{\kappa \over (\kappa + 2)}} y^{-\kappa \over
(\kappa + 2)}
\end{eqnarray}
Using these expressions we can relate $c$ and $\kappa$ with $a$
and $\omega$:
\begin{equation}
\kappa   =  { 2\omega \over ( 1 - \omega)}, (kc)^{ 1/ ( k + 2)}
=(2 a \omega)^{1/2}, (1 - \omega)a = k^{-\omega} c^{1 - \omega}.
\end{equation}
It is now possible to express the asymptotic form of $\mu_n$ as a
function of $y$ and in terms of parameters describing the
asymptotic behaviour of $\sigma(N_c)$, i.e. $a$ and $\omega$:
\begin{eqnarray}
\mu_1(y) & = & \tau(y) = (2a \omega)^{1/2} y^{(1 -
\omega)/2}\nonumber
\\ \mu_2(y) & = & {1 + \omega \over 2}\nonumber  \\
\mu_3(y) & = & { 1 - \omega^2 \over 4 }  (2 \omega a)^{-1/2}
y^{-(1 - \omega)/2}\nonumber \\ \mu_4(y) & = & { ( 1 - \omega^2
)(6 + 2\omega) \over 8 } (2 \omega a)^{-1} y^{ -( 1 - \omega)}.
\end{eqnarray}
Using these asymptotic forms and the definitions of
$h(>x)\mu_n(>x)$ introduced earlier in equations (28), we can can
write
\begin{eqnarray}
h(>x) & = & a {x^{\omega-1}\over \Gamma(\omega)}\nonumber  \\
\nu_1(>x)h(>x) & = & h(>x)\mu_1(>x) = (2 \omega
a)^{1/2}{x^{(\omega -1)/2} \over \Gamma \big [ {1 + \omega \over
2} \big ]} \nonumber\\ \nu_2(>x)h(>x) & = & {1 + \omega \over 2}
\nonumber \\ \nu_3(>x)h(>x) & = & \Big ( { 1 - \omega^2 \over 4 }
\Big ) (2 \omega a)^{-1/2} {x^{(1 - \omega)/2} \over \Gamma \big [
{ 3 - \omega \over 2} \big ] } \nonumber\\ \nu_4(>x)h(>x) & = & {
( 1 - \omega^2 )(6 + 2\omega) \over 64 \omega^3}(2 \omega
a)^{-1}{x^{2(1 - \omega)} \over \Gamma \big [ { 3 - \omega \over
2} \big ] }.
\end{eqnarray}
It is interesting to note the power of the bias $b(>x)$ that
appears in each of these expressions is precisely that required
that, when divided by the appropriate quantity  to evaluate $M_s$,
the parameters for collapsed objects in the limit $x<<1$  all
become independent of $x$:
\begin{eqnarray}
M_3(x) & = & { ( 1 + \omega) \over 4} { \Gamma \big [{ ( 1 +
\omega ) \over 2} \big ] \over \Gamma[{1+\omega}]};\nonumber \\
M_4(x) & =  & { ( 1 - \omega^2 ) \over 16 \omega^2} { \Gamma \big
[{ 1 + \omega \over 2} \big ]^3 \over \Gamma[\omega]^2
\Gamma[{3-\omega\over 2}]};\nonumber \\ M_5(x) & = & { ( 1 -
\omega^2 )(6 + 2\omega) \over 64 \omega^3} { \Gamma \big [{ 1 +
\omega \over 2} \big ]^4 \over \Gamma[\omega]^3
\Gamma[{2-\omega}]}.
\end{eqnarray}
The implications for cumulants and cumulant correlators of
overdense regions are illustrated in Figure 3.

\begin{figure}
\protect\centerline{
\epsfysize = 2.85truein \epsfbox[18 434 589 718]
{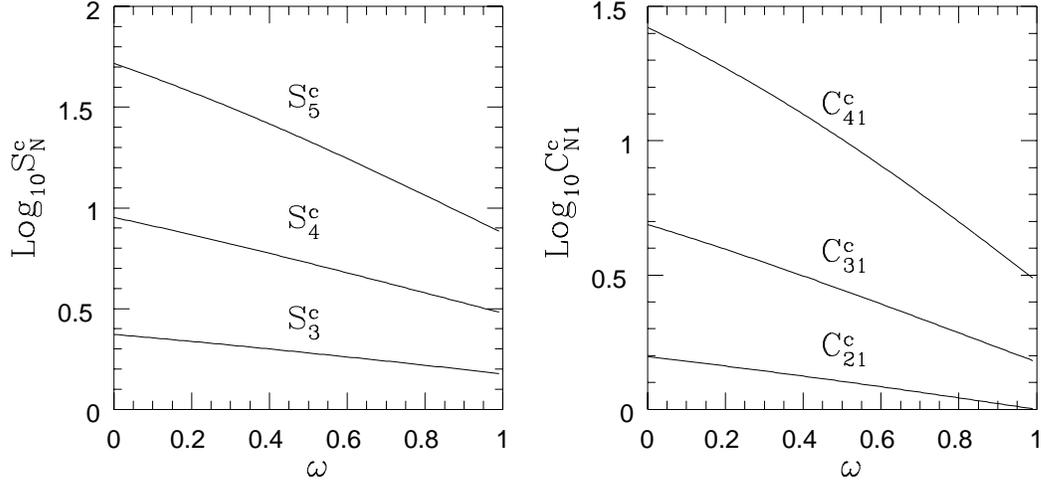} } \caption{Lower-order cumulants and cumulant
correlators are plotted as a function of $\omega$ for the limiting
case of $x<<1$. There are no analytical results connecting
$\omega$ with the initial spectral index, $n$, but simulations
suggest that $\omega$ increases with the amount of small-scale
power. Determinations of $\omega$ for different scale free initial
conditions in 2D and 3D were made by Munshi et al. (1997). Our
results show that both cumulants and cumulant correlators for
overdense regions increase with the amount of small-scale power,
which reflects similar a behaviour for matter cumulants and
cumulant correlators. }
\end{figure}

\subsection{The case of highly over-dense objects ($x >>1$)}
It is also possible to evaluate the $S_N$ parameters in the limit
when $x >>1$. The dominant contribution to this case comes from
the singular nature of $\sigma(N_c)$ for small negative values of
$y$. The value of $y = y_s$ for which $\sigma(y)$ becomes singular
is given by the solution of the equation
\begin{equation}
y_s = -{\tau_s \over G'(\tau_s))},
\end{equation}
where $\tau_s$ can be evaluated by solving the implicit equation
\begin{equation}
\tau_s = { G'(\tau_s) \over G''(\tau_s)}
\end{equation}
which really depends on the exact modelling of $G(\tau)$ in the
nonlinear regime. However, as we will show, in the limit of rare
events where $x$ takes very large values, the hierarchical
amplitudes become independent of the initial conditions and this
considerably simplifies things. The quantity $\sigma_(y)$ plays a
very important role in determining various scaling properties and
is known to have a singularity for small but negative values of
$y$ which determines the asymptotic values of $S_N$ for large
values of $x$. We can express $\sigma(y)$ near the singularity as
\begin{equation}
\sigma - \sigma_s = -a_s \Gamma \big( -{3 \over 2} \big) ( y -
y_s)^{3/2}; \end{equation} with \begin{equation} a_s = {1 \over {
\Gamma \big( -{1 \over 2} \big)}} G'(\tau_s)G''(\tau_s) \left( { 2
G'(\tau_s) G''(\tau_s) \over G'''(\tau_s)} \right).
\end{equation}
Such a singularity  produces an exponential cut-off in the power
law profile of  $h(x)$ above, and which we can express as
\begin{equation}
h(x) = -a_a y_s x^{-5/2} \exp(y_sx).
\end{equation}
Differentiating $y$ twice with respect to $x$ and using the fact
that the first order derivative of $\tau$ with respect to $y$
vanishes at the singular point $y_s$ it is possible to show that,
near the singular point,
\begin{equation}
(\tau - \tau_s) = \Big ( { 2 G'(\tau_s) G''(\tau_s) \over G'''(\tau_s)} \Big )^{1 /2}
( y - y_s)^{1 /2}.
\end{equation}
Similarly, the other $\mu_n(y)$ functions can also be expressed as
a function of $y$ near the singularity $y_s$. We list only those
terms which will dominate in determining the $S_N$ parameters for
large values of $x$.
\begin{eqnarray}
\mu_2(y) & = &{ -y G''(\tau) \over 1 + yG''(\tau)} = { G''(\tau_s)
\over G'''(\tau_s)} \Big ( {2G'(\tau_s) G''(\tau_s) \over
G'''(\tau_s)} \Big )^{-1/2} (y - y_s)^{-1/2} \nonumber\\ \mu_3(y)
& = & { -y G'''(\tau) \over (1 + yG''(\tau))^3} = { G''(\tau_s)^2
\over G''(\tau_s)} \Big ( {2G'(\tau_s) G''(\tau_s) \over
G'''(\tau_s)^3} \Big )^{-3/2} (y - y_s)^{-3/2}
 \nonumber\\
\mu_4(y) & = & { -y G^{(IV)}(\tau) \over ( 1 + yG''(\tau))^4} + {
3 y^2 G'''(\tau)^2 \over ( 1 + yG''(\tau))^5} = { 3G''(\tau_s)^3
\over G'''(\tau_s)^3 G'(\tau_s)}\Big ( {2G'(\tau_s) G''(\tau_s)
\over G'''(\tau_s)}\Big )^{-5/2} (y - y_s)^{ -5/2}.
\end{eqnarray}
Using these relations in the expressions for $\mu_n(>x)$, it is
possible to show that,
\begin{eqnarray}
b(>x) & = & \mu_1(>x) = -\Big ( { x \over G'(\tau_s)} \Big
)\nonumber\\ \mu_2(>x) & = & \Big ( { x \over G'(\tau_s)} \Big
)^2\\ \mu_3(>x) & = & - \Big ( { x \over G'(\tau_s)} \Big )^3
\nonumber \\ \mu_4(>x) & = & \Big ( { x \over G'(\tau_s)} \Big
)^4.
\end{eqnarray}
Using the appropriate definitions of the $S_N$ and $C_{NM}$
parameters, it is now straightforward to obtain
\begin{equation}
S_N^c = N^{N - 2},~~~ C_{N1}^c = N^{N - 1},~~~~ C_{N11}^c = N^N.
\end{equation}

It is interesting that all the $\mu_n(>x)$ become equal to unity
in this limit, so that the $S_N$ parameters are simply equal to
the number of tree diagrams of that order, i.e. $N^{N-2}$. Exactly
the same result has also been derived from the extended
Press-Schechter formalism (Press \& Schechter 1974; Mo et al. 96).
It is remarkable and encouraging that the hierarchical ansatz,
which is known to be valid only in the highly non-linear regime,
predicts the same results as this alternative formalism, which is
based on smoothing of the initial  density field.

\section{Discussion}

Our starting point for this study has been the  most general form
of hierarchical model for the matter correlation functions in the
highly non-linear regime. This is equivalent to the assumption
that the correlation hierarchy can be built up as  a tree
structure where each vertex is assigned a different weight (the
vertex amplitude), but which is independent of geometrical form
factors. The amplitudes associated with vertices are left
arbitrary. We have adopted the method of generating functions
introduced by Bernardeau \& Schaeffer (1992) to show that such a
hierarchy induces a similar tree structure for overdense cells,
which we take to represent {\em collapsed objects}.

We then expanded the generating functions for the joint
probability distributions of several cells  in terms of a series
in powers of $(\xi_{ij}/ \bar \xi_i)$.  At zeroth order it
reproduces one-point PDF with no correlation between different
cells. At linear order of $(\xi_{ij}/ \bar \xi_i)$ two-point
correlations between different cells can be used to compute the
bias of overdense cells with respect to the underlying matter
distribution. At second order of $(\xi_ij/ \bar \xi_i)$ a
hierarchical form for the three-point correlation of overdense
cells is reproduced and this can be used to compute  the
hierarchical amplitude $M_3$ associated with collapsed objects.
Similarly, each higher-order term reproduces the amplitude of a
new vertex and the associated multipoint cumulant correlators
(MCCs) for collapsed objects. Our analysis is also able to
reproduce the exact number of all the higher-order hybrid and
snake diagrams with their associated tree amplitudes. We found
that the one-point and the multi-point statistics of collapsed
objects depend physically on properties intrinsic to the objects,
and thus depend mathematically only on the scaling variable $x =
N/N_c$.

Assuming a very general model for the generating function
$G(\tau)$, for the underlying mass we have computed cumulants
$S_N^c$ and cumulant correlators $C_{N1}^c$ for overdense cells.
We have carried out this analysis in two limiting cases, of
moderately overdense cells $x<<1$ and extremely overdense cells
$x>>1$, respectively. We find that for extremely overdense cells
the results are independent of initial conditions, the amplitudes
associated with vertices of arbitrary order tend asymptotically to
unity, and the generating functions can be expressed in a closed
form $G(\tau) = \exp(-\tau)$. This closed-form solution can be
used to compute all other statistical quantities for over-dense
cells such as the count probability distribution function (CPDF)
or the void probability function (VPF) for overdense cells. For
moderately overdense cells, the results depend on the initial
power spectrum and we find that the $S_N$ parameters for overdense
cells decrease with increasing small-scale power in this case.

We have also developed a diagrammatic method for counting
different ways in which a general correlation function $\xi_N$
with $N$ arguments can be grouped in such a way that $p$ of its
nodes are associated with one particular cell, $q$ are in a second
cell, and so on. In leading order such a grouping of vertices in
different cells generates a new level of the tree hierarchy which
can be related to the correlation hierarchy for overdense cells.
The only difference between the trees for the underlying mass
distribution and those for collapsed objects lies in the internal
structure of their constituent nodes. Trees associated with
collapsed objects have internal degrees of freedom for different
vertices that appear in different ways. As we have shown, however,
it is possible to group nodes associated with underlying mass
trees in such a way as to replicate the tree structure for
collapsed objects with same external topology but different
internal weights. Our method of counting internal degrees of
freedom associated with the vertices matches with analytical
results obtained by more laborious means.

In an exactly similar manner it is possible to use the trees
associated with collapsed objects to study how clusters or groups
of collapsed objects are correlated with each other. In this case,
the nodes associated with groups of collapsed objects will have
many more internal degrees of freedom as they will consist of
nodes of tree structures of collapsed objects which themselves are
made of point-like vertices of the underlying mass distribution.
Amplitudes associated with nodes appearing in the tree structure
of groups of collapsed objects will definitely  depend on the mass
of  objects that constitute the group, and will have scaling
properties associated with them. We shall present a complete
analysis of this problem in near future.

>From an observational point of view, the results contained in this
paper are important for estimating finite volume corrections. It
is well known that the finite size of galaxy catalogs introduce
errors in estimations of cumulants and cumulant correlators. Error
estimation for values of the $S_N$ parameters --  normalised
moments of the CPDF -- extracted from finite catalogs involve
volume integrals of the two-point joint probability distribution
function. Such an analysis, carried out by Szapudi \& Colombi
(1996),  involves the generating function $\mu_1(y) = \tau(y)$.
Similar estimation procedures for multi-point cumulant correlators
will involve higher-order generating functions describing
correlation between more than two cells. Our analytical
expressions for $\mu_N$ will be useful in such analyses. In a
slightly different approach, it is also possible to talk about the
probability distribution in the context of error estimation.
Estimated values of different statistical quantities are affected
by the presence or absence of rare objects in a finite size
catalog. The higher the order of cumulant, the higher the
contribution from the high-$N$ tail of the CPDF, which is highly
sensitive to presence or absence of rare objects. It is natural,
therefore, to expect that our analytical results determining the
generating function for rare objects can be related to the
probability distribution of estimated errors. This will extend
earlier studies in this direction where only the variance of
errors were considered. This is important because we know that
finite volume correction not only introduces large error but
measurements are more likely to underestimate the true values of
cumulants or cumulant correlators.

Cumulants for collapsed objects have also been computed using
alternative procedures, chiefly by extension of the
Press-Schechter formalism (Bond et al. 1991; Bower 1992). Our
results complement these studies, because they are based on a
fully nonlinear approach and depend only on the assumption that
there is a  hierarchical structure of the correlation functions.
We think this {\em ansatz} is based in relatively firm ground, and
can, at least in principle, be directly related to the microscopic
physics of gravitational clustering through the BBGKY equations.
The analysis can also be applied the MCCs of arbitrary order which
have not yet been investigated using any other formalism. Our
results for computation of $S_N$ parameters for collapsed objects
match with such an analysis done using extended Press-Schechter
formalism. However similar comparison for moderately over-dense
objects are difficult as different prescriptions are used to
define collapsed objects in these two formalisms. We plan to
compare our theoretical predictions against numerical simulations
both in 3D as well as in 2D, and will present these results in due
course.

It is also important to note that, owing to our incomplete
knowledge of gravitational clustering in the highly non-linear
regime, at least two different {\em ansatze} are often used to
study nonlinear clustering, namely the  hierarchical ansatz (which
we discuss here)  and the stable clustering {\em ansatz} and these
should not be confused. While the hierarchical {\em ansatz}
relates higher-order correlation functions with the two-point
correlation function, the stable clustering {\em ansatz} on the
other hand relates the power-law index of the two-point
correlation with the initial power spectral index. Thus these two
{\em ansatze} are not contradictory but complement each other. Our
analysis is based completely on the hierarchical {\em ansatz}, and
we have never used stable clustering  to derive our results.

It may also be possible to extend our analysis to the joint
probability distribution function of densities of concentric
cells, with their centres coinciding with the centre of a dark
halo. In such a context it will be interesting to check if our
analysis can also predict the inner structure of haloes , i.e.
their density profiles. This will also clarify the issue of
whether there exists a  universal halo profile as proposed by
Navarro et al. (1996, 1997) and how it relates to stable
clustering and to the hierarchical ansatz. We will report these
results elsewhere.

\section*{Acknowledgments}
Dipak Munshi acknowledges support from PPARC under the QMW
Astronomy Rolling Grant GR/K94133. Peter Coles received a PPARC
Advanced Research Fellowship during the period when most of this
work was completed. We are grateful for support under the
NSF-EPSCoR program, as well as the Visiting Professorship and
General Funds of the University of Kansas.

\end{document}